\documentstyle[12pt]{article}
\newcommand{\lsim}{{\, \lower2truept\hbox{
${< \atop\hbox{\raise4truept\hbox{$\sim$}}}$}\,}}
\newcommand{\gsim}{{\, \lower2truept\hbox{
${> \atop\hbox{\raise4truept\hbox{$\sim$}}}$}\,}}

\begin{document}
\baselineskip=18pt

$\!$

\centerline{\Large\bf Rebirth of Novae as Distance Indicators}
\centerline{\Large\bf Due to Efficient Large Telescopes} 
\bigskip

\vskip 1.5truecm
\centerline {\sl Massimo Della Valle$^{1}$}
\vskip .4cm
\centerline {\sl Roberto Gilmozzi$^{2}$}
\vskip 1.5truecm 

\centerline{$^1$Osservatorio Astrofisico di Arcetri, Firenze, Italy}
\centerline{$^2$European Southern Observatory, Paranal, Antofagasta, Chile} 
\vskip 1.5truecm 

\centerline{{\small in 17 May 2002 Science issue}}
\vskip 1.5truecm

A nova is a close binary system, where one
component is a white dwarf. A nova exhibits a sudden and rapid
increase in its brightness because of thermonuclear reactions on the
surface of the white dwarf that is accreting hydrogen-rich material
from its smaller mass companion star.These explosions liberate about
10$^{45}$ ergs of energy within a few weeks, thus making novae some of the
most luminous transient sources in the sky and, therefore, powerful
standard candles for measuring intergalactic distances (1,2).  In
addition nova surveys in external galaxies can be used to determine
the average number of nova outbursts per year, the nova rate, and this
rate can be used to estimate the contribution of novae to the chemical
evolution of the parent galaxy (3) and their potential to be gamma-ray
producers (4) . Despite the importance of novae, they are difficult to
detect and observe in external galaxies with 2 to 4 m class
telescopes. Here we used the 8.2 meter Very Large Telescope (VLT) to
search for novae in NGC 1316, the parent galaxy of the type Ia
Supernovae 1980N and 1981D.  The observations were performed during
nine nights between 25 December 1999 to 19 January 2000. They were
carried out in service mode at the VLT equipped with the FORS-1
instrument (focal reducer/low dispersion spectrograph) and a 2048 by
2048 charged coupled device (CCD) camera with a projected pixel size
of 0.2 arc seconds and a field of view of 6.8 x 6.8 arc minutes. Each
20-minute exposure was imaged with filter B (in the Bessel photometric
system), sometimes complemented by V and I.  The background light due
to the galaxy was removed by applying a median filter to each image,
which was successively subtracted from the original frame. This
procedure generates images containing only stars and the faint galaxies.
The novae were discovered by comparing each -background-subtracted- B
frame, with the one obtained on 25 December. Photometric measurements
have been performed with Sextractor(5) and the aperture photometry was
corrected to account for seeing variations. We found 4 transient
objects (Fig. 1 ) with blue colors, (B-V) ~0 , which are typical for
novae observed around the maximum of their brightness. The time scale
of the apparent brightness and color variability of these objects
(Fig. 2) are inconsistent with other types of variable stars, such as
Mira, cepheids, Hubble-Sandage variables or foreground objects like
RR Lyr and flare stars. These four novae in NGC 1316, at a distance of
$\sim 19.5$ Mpc, are the first detected beyond the Virgo cluster. The
sampling of the lightcurves is adequate to estimate the distance to
the galaxy with the Buscombe-de Vaucouleurs relation. Because the
last data points of the light curves were obtained within 20 days of
the brightness maximum, the corresponding apparent magnitudes allow us
to set an upper limit to the distance of the galaxy of 20 Mpc $\pm 2.4$ 
(1$\sigma$). Nova A was caught during its early decline from maximum
brightness, therefore the last data point can only be used to set a
lower limit to the distance, i.e.  18.2 Mpc $\pm 2.2$. The estimated
distances imply an absolute magnitude at maximum brightness of M$_B \gsim
-19.20\pm 0.35$ and M$_B \gsim -19.10\pm 0.35$ for SN 1980N and SN 1981D,
respectively. This result is consistent with the existence of a $\sim 0.2$
to $0.3$ mag deficiency in the luminosity at maximum of type Ia
Supernovae found in early type galaxies compared to SNe found in
spirals (6). Simulated VLT observations of novae in the Fornax
cluster (7), where NGC 1316 is located, showed that our novae sample
might be incomplete by as much as 20\%. With this in mind and by
applying the control time technique (8), we estimate a nova rate for
NGC 1316 of about 90 to 180 novae per year. After normalizing this
rate to the infrared luminosity of the galaxy, we find that NGC 1316
tends to produce novae less prolifically than some type of spiral
galaxies (9,10). Novae can be used as distance indicators like
cepheids by studying the Zwicky and Buscombe-de Vaucouleurs
relationships in parent galaxies with well observed type Ia SNe. Novae
that can be rapidly and easily detected with larger telescopes, such
as the VLT have another important advantage over cepheids because they
can be observed in all type galaxies.  
\bigskip

References 

1. van den Bergh,S., Pritchet, C.J. 1986, PASP, 98, 110 

2. Della Valle, M., Livio, M. 1995, ApJ, 452, 704 

3. Gehrz, R.D., Truran, J.W., Williams, R.E.,Starrfield, S. 1998, PASP, 110,3 

4. Gomez-Gomar, J., Hernanz, M., Jose, J., Isern, J. 1998, MNRAS, 296, 913 

5. Bertin, E., Arnouts, S. 1996, A\&A Suppl. Ser. 117, 393 

6. Branch, D., Romanishin, W., Baron, E. 1996, ApJ, 465, 73 

7. Della Valle, M., Gilmozzi, R. 1996, in 'The Early Universe with VLT', J. Bergeron Ed. (Springer, Berlin 1997), p. 380  

8. Zwicky, F. 1942, ApJ, 96, 28 

9. Della Valle, M., Rosino, L., Bianchini, A., Livio, M. 1994, A\&A, 287, 403 

10. Yungelson, L., Livio, M., Tutukov, A. 1997, ApJ, 481, 127

\newpage  

Fig.1.  Novae in B light during the bright and faint states: nova A (9
and 13 Jan 2001); nova B (26 Dec 2000 and 12 Jan 2001); nova C (26 Dec
2000 and 12 Jan 2001 ); nova D (9 and 12 Jan 2001).
\bigskip
\bigskip

Fig. 2. Photometric evolution of the Novae.  Filled Circles: observed
B magnitudes; filled triangles: observed V magnitudes; filled squares:
observed I magnitudes. v symbols represent upper limits.
 
\end{document}